\documentclass[aps,pre,showpacs,amsmath,amssymb,amsfonts,lengthcheck,twocolumn,superscriptaddress]{revtex4}

\usepackage{changes} 

\usepackage{graphicx}
\usepackage{subfigure}
\usepackage{eucal}
\usepackage{setspace}
\usepackage{pifont}
\usepackage{verbatim}
\usepackage{dcolumn}
\usepackage{bm}
\usepackage{color}
\usepackage[colorlinks=true,citecolor=blue,linkcolor=blue,urlcolor=blue]{hyperref}%
\usepackage{xcolor}
\usepackage{dsfont}
\usepackage{longtable}
\usepackage{hyperref}
\usepackage{multirow}
\usepackage{xcolor}
\usepackage{amsmath}
\usepackage{amssymb}
\allowdisplaybreaks
\usepackage[utf8]{inputenc}   
\usepackage[T1]{fontenc}
\allowdisplaybreaks

\newlength{\wdth}

\newcommand{\bla}{bla\\bla\\bla\\bla\\bla}

\begin{document}

\title{Entropy–Seebeck Ratio as a Tool for Elementary Charge Determination}


\author{Francisco J. Peña}
\affiliation{Departamento de Física, Universidad Técnica Federico Santa María, 2390123 Valparaíso, Chile}

\author{César D. Nuñez}
\affiliation{Departamento de Física, Universidad Técnica Federico Santa María, 2390123 Valparaíso, Chile}

\author{Bastian Castorene}
\affiliation{Instituto de Física, Pontificia Universidad Católica de Valparaíso, Casilla 4950, 2373223 Valparaíso,
Chile}
\affiliation{Departamento de Física, Universidad Técnica Federico Santa María, 2390123 Valparaíso, Chile}

\author{Michel Aguilera}
\affiliation{Instituto de Física, Pontificia Universidad Católica de Valparaíso, Casilla 4950, 2373223 Valparaíso,
Chile}
\affiliation{Departamento de Física, Universidad Técnica Federico Santa María, 2390123 Valparaíso, Chile}

\author{Natalia Cortés}
\email{natalia.cortesm@usm.cl}
\affiliation{Departamento de Física, Universidad Técnica Federico Santa María, 2390123 Valparaíso, Chile}

\author{Patricio Vargas}
\affiliation{Departamento de Física, Universidad Técnica Federico Santa María, 2390123 Valparaíso, Chile}

\date{\today}

\begin{abstract}
In this work, we investigate the relationship between the Seebeck coefficient $(\mathcal{S})$, and the differential entropy per particle (DEP, $s$), as a tool for characterizing charge carriers in two-dimensional systems. Using armchair silicene nanoribbons as a model platform, we analyze how both quantities and their ratio depend on chemical potential at room temperature. While the Seebeck coefficient captures transport properties through the energy dependence of the electronic transmission, the DEP is directly connected to the system's electronic entropy, offering a direct thermodynamic alternative for estimating $\mathcal{S}$. We evaluate these transport-thermodynamic properties considering diverse ribbon widths, defining metallic and semiconducting regimes. We find both quantities $\mathcal{S}$ and $s$, are highly interconnected within the ribbon's band gap energy region, and their ratio $s/\mathcal{S}$ converges to the elementary charge $e$ across that energy window, fulfilling the Kelvin formula $\mathcal{S}=s/e$. On the contrary, $s/\mathcal{S}$ is undefined for gapless ribbons in the energy window of the first transmission channel.
These results establish the ratio between the DEP and the Seebeck coefficient as a reliable and complementary probe for the determination of the elementary charge, and to identify the cleanness of electronic band gaps as $s/\mathcal{S}$ matches with $e$.     
\end{abstract}

\maketitle


\section{\label{sec:level1} Introduction}

The precise determination of the elementary charge \(e\) is a foundational aspect of electromagnetism, with implications that span from fundamental metrology to the advancement of quantum technologies~\cite{metro1,metro2,poirier2019ampere}. Since Millikan’s oil-drop experiment~\cite{Millikan1913}, other techniques including the counting of electrons on a capacitor \cite{international2001international},
and single-electron charge sensing in quantum dots~\cite{kiyama2018single}, have expanded the possibility to determine and improve the accuracy of measurement of $e$.

Despite these advances, a direct extraction of a pure value of $e$ remains challenging in nanoscale systems, as interaction effects often mask the plain microscopic charge content, complicating its identification \cite{vanevic2008elementary}.
%
However, a different approach from the connection of thermoelectrics and thermodynamics observables can be implemented, offering a compelling framework for exploring this problem. 

Thermoelectric effects quantified as the Seebeck coefficient  $\mathcal{S}$, which measures the voltage drop in response to a temperature gradient, can be expressed within the linear response regime in terms of energy integrals of the electronic transmission function~\cite{tritt2011thermoelectric,goupil2011thermodynamics,alam2013review,liu2012recent}. At low temperatures using the Sommerfeld expansion, $\mathcal{S}$ converts in the Mott formula~\cite{Cutler1969Mott}
\begin{equation}
\mathcal{S}(\mu,T) \;=\; -\frac{\pi^2}{3} \frac{k_B^2 T}{e} \left.\frac{d \ln \sigma(\epsilon)}{d\epsilon}\right|_{\epsilon=\mu},
\label{eq:Mott}
\end{equation}
which captures a linear dependence on temperature $T$, and the role of the energy derivative of the electronic conductivity $\sigma(\epsilon)$. In the opposite regime, the high-temperature limit yields the Mott-Heikes expression
\begin{equation}
\mathcal{S}(\mu,T) \;=\; -\frac{\mu}{eT},
\label{eq:MottHeikes}
\end{equation}
which becomes exact in many correlated insulators~\cite{Heikes1961,Jones1956}. These formulations for $\mathcal{S}$ in Eq.\ \ref{eq:Mott} and Eq.\ \ref{eq:MottHeikes} underscores the sensitivity of particle--hole asymmetries around the chemical potential \(\mu\) of electronic systems, and partially give us information of $e$ as included in their denominators.

A thermodynamic quantity intimately related to the Seebeck coefficient is the differential entropy per particle (DEP), $s$,
\begin{equation}
\label{eq:Sp_def}
s(\mu,T)=
\left(\frac{\partial S}{\partial N}\right)_{T,V}
=-\left(\frac{\partial\mu}{\partial T}\right)_{N,V},
\end{equation}
where \(S\) is the total electronic entropy, and \(N\) the total number of electrons in the system. This thermodynamic quantity possesses high sensitivity in electronic measurements of two-dimensional (2D) strongly correlated systems~\cite{kuntsevich2015strongly}, and it has been theoretically obtained for different 2D materials, including germanene, gapped graphene, semiconducting dichalcogenides, zigzag graphene ribbons, and graphene under magnetic and electric fields \cite{Chaika2025, cortes2023entropy, Shubnyi2018}.

Additionally to its intrinsic significance because of the direct link with the electronic entropy $S$, the definition for $s$ also provides a direct link to thermal electronic transport: it naturally follows from Kelvin’s thermodynamic formulation of the Seebeck coefficient ~\cite{peterson2010kelvin,kokalj2015enhancement,Yamamoto2017MagKelvin}
\begin{equation}\label{eq:Kelvin}
    \mathcal{S}(\mu,T)=\frac{1}{e}\left(\frac{\partial S}{\partial N}\right)_{T,V}.
\end{equation}
%
The comparison of Eq.~\eqref{eq:Sp_def} and Eq.~\eqref{eq:Kelvin} generates an explicit correspondence between $s$ and $\mathcal{S}$, telling us that one can compute/measure a pure value of $e$ through the ratio
\begin{equation}
\label{eq:charge_relation}
\frac{s(\mu,T)}{\mathcal{S}(\mu,T)}=e.
\end{equation}
%
%
%
Equation \ref{eq:charge_relation} holds whenever the spectral weights involved in \(s\) and \(\mathcal{S}\) are proportional or vanish within the same energy window, the latter occurring in gapped zigzag graphene ribbons \cite{cortes2023entropy}. 
Moreover, Eq.~\eqref{eq:charge_relation} offers a thermodynamic route as $s=e \mathcal{S}$. When this relationship is fulfilled, the Seebeck coefficient $\mathcal{S}$ can be identified as the transported differential entropy per charge, therefore, avoiding the need for explicit electronic transport calculations. 


To test the validity and precision of Eq.~\eqref{eq:charge_relation}, we use armchair silicene nanoribbons (ASiNRs) because their electronic properties alternate between gapped and gapless behavior as a function of the electron allowed energies when the ribbon's width $N$ changes, providing controlled access to both regimes around $\mu=0$ eV ~\cite{chen2019thermoelectrics,kim2013thermoelectricity,cortes2016enhancement,dominguez2019nanowires,WOS:000690816500001,WOS:000365365300046,WOS:000371393800007}. We also identify the spectral regimes in which the DEP closely follows the behavior of the Seebeck coefficient. We find there is a high correspondence $s \simeq \mathcal{S}$ with clear peak-dip line shapes for both quantities within the band gap energy region of semiconductor ASiNRs, while in gapless ribbons $\mathcal{S}$ and $s$ show deviations between them in the energy window of the ribbon's first transmission channel. 

By computing the ratio $s/\mathcal{S}$ for gapped ribbons, we access to a direct estimation of $e$ and its fluctuations in a wide energy range ($\mu > 1$ eV) for ribbons with width $N=10$ and $N=12$. These results show that $s/\mathcal{S}$ is mirror symmetric about $\mu=0$ eV, fluctuating between negative and positive values outside the band gap energy region, and stabilizing within it with a near flat plateau of value $\simeq e$. The ratio linearly changes with temperature within the plateau for a fixed value of $\mu \neq 0$. All these responses demonstrate that $s/\mathcal{S}$ provides a direct estimate of the elementary charge at/around room temperature. 


\section{\label{sec:model}System and Model}


\begin{figure}[b]  
  \centering
\includegraphics[width=\columnwidth]{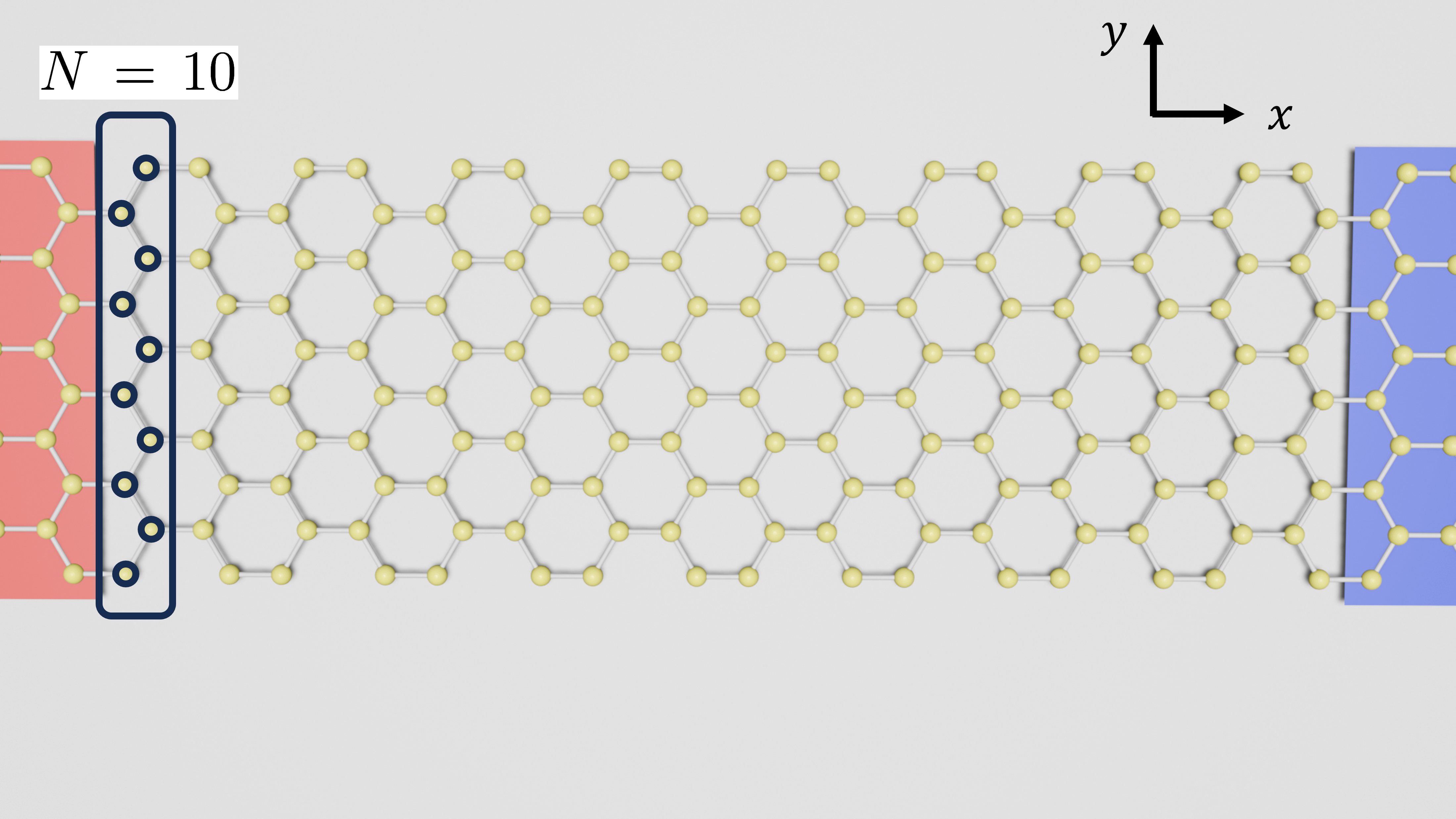}
 \caption{Schematic view of an armchair silicene nanoribbon oriented along the \(x\) axis. The black rectangle indicates the ribbon unit cell, which contains \(N = 10\) atoms across its width, and yellow spheres represent silicon atoms. The left (right) red (blue) region shows the semi-infinite leads producing a thermal gradient bias.}
  \label{Fig.sketch}
\end{figure}

\begin{figure*}[t]
  \centering
  \includegraphics[width=14.7cm]{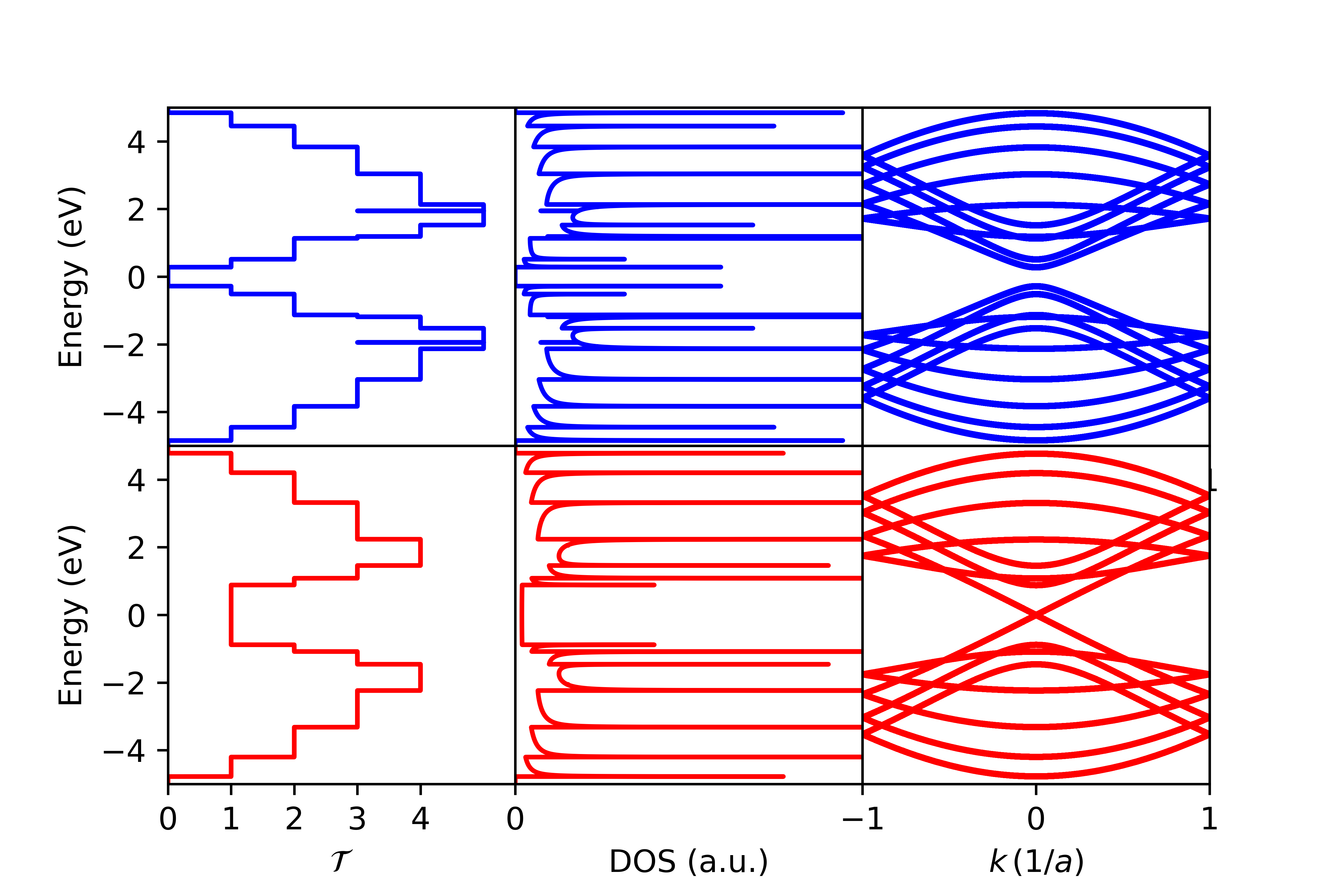}
  \caption{Energy-dependent transmission $\mathcal{T(\varepsilon)}$, density of states (DOS) in arbitrary units, and electronic band structure as a function of momentum $k$ ($a$ is the silicene lattice constant)
 for two armchair silicene nanoribbons. The top panel (blue lines) corresponds to a semiconducting ribbon with $N=10$, while the bottom panel (red lines) shows a metallic case with $N=8$. The energy range is limited to $|\varepsilon| \le 5$\,eV, which contains all the relevant ribbon's bands.}
  \label{Fig.TDB}
\end{figure*}

The systems under study are rectangular armchair silicene nanoribbons, electrically contacted to semi-infinite leads of the same material, as schematically illustrated in Fig.~\ref{Fig.sketch} for a ribbon with width $N=10$.  The electronic states in the central region are described by a single
$\pi$‑orbital tight‑binding Hamiltonian~\cite{Ezawa2012}

\begin{equation}
  \label{eq:Hamiltonian}
  H_C = \sum_{i} \varepsilon_i c_{i}^{\dagger} c_{i}
        - t \sum_{\langle i,j \rangle} \left( c_{i}^{\dagger} c_{j} + \text{h.c.} \right),
\end{equation}
where  $c^{\dagger}_{i}$ ($c_{i}$) creates (annihilates) an electron at the site $i$ and the sum runs over all nearest-neighbor pairs $\langle i,j \rangle$. The first term corresponds to the on-site energy $\varepsilon_i$ at site $i$ and the second term represents the nearest-neighbor hopping with amplitude $t = 1.6$ eV. For simplicity, we have fixed the on-site energy at $\varepsilon_i=0$ throughout this work.

Electronic transport is treated within the non-equilibrium Green's function formalism combined with decimation techniques~\cite{lambert1980decimation}. In the linear response regime, the transmission function, $\mathcal{T}(\varepsilon)$, is given by~\cite{Datta_1995}

\begin{equation}
  \label{eq:Transmission}
  \mathcal{T}(\varepsilon) =
    \mathrm{Tr}\!\big[\Gamma_L(\varepsilon)\,
    \mathcal{G}^{\dagger}(\varepsilon)\,
    \Gamma_R(\varepsilon)\,
    \mathcal{G}(\varepsilon)\big].
\end{equation}

The retarded Green's function is given by \(\mathcal{G}(\varepsilon) = \big[\varepsilon - H_C - \Sigma_L(\varepsilon) - \Sigma_R(\varepsilon)\big]^{-1}\), with $\varepsilon$ the allowed electron energies. The linewidth matrices \(\Gamma_{\alpha}(\varepsilon) = i\big[\Sigma_{\alpha}(\varepsilon) - \Sigma_{\alpha}^{\dagger}(\varepsilon)\big]\) (\(\alpha = L, R\)) incorporate the coupling to the leads, whose self-energies are defined as \(\Sigma_{\alpha}(\varepsilon) = V_{C\alpha}\,g_{\alpha}(\varepsilon)\,V_{\alpha C}\). Throughout, all energy-dependent quantities are calculated under the prescription
\(\varepsilon \to \varepsilon + i\eta\) and \(\eta \to 0^+\).

The density of states (DOS), \(D(\varepsilon)\), is given by
\begin{equation}
  \label{eq:DOS}
  D(\varepsilon) = -\frac{1}{\pi}\,
  \lim_{\eta \to 0^+}
  \mathrm{Tr}\!\left[ \mathrm{Im}\,\mathcal{G}(\varepsilon + i\eta) \right].
\end{equation}

For armchair nanoribbons, the electronic character depends on the number of atoms \(N\) across the width. When \(N = 3p + 2\) (\(p \in \mathbb{Z}\)), the ribbon is metallic; otherwise (\(N = 3p\) or \(N = 3p + 1\)), it is a semiconductor. The example shown in Fig.~\ref{Fig.sketch} corresponds to \(N = 10\), i.e., a semiconducting armchair silicene nanoribbon (10-ASiNR).

Fig.~\ref{Fig.TDB} compares the transmission, DOS, and band dispersion for a metallic 8-ASiNR (red lines, lower panels), and a semiconducting 10-ASiNR (blue lines, upper panels). For the metallic 8-ASiNR, the valence and conduction bands show linear dispersion about zero energy, resulting in a finite DOS and a single transmission channel about \(\varepsilon = 0\). In contrast, the 10-ASiNR exhibits a bandgap around zero energy, leading to a nearly vanishing DOS and transmission within that energy region. 

In what follows, we use these electronic transport and DOS results, and different widths for the energy spectra of ASiNR, to calculate the thermoelectric (Seebeck coefficient, $\mathcal{S}$) and thermodynamic (DEP, $s$) responses, as well as their ratio $s/\mathcal{S}$.      
 
\section{\label{sec:level3} Seebeck Coefficient}

To characterize the thermoelectric response of armchair silicene nanoribbons, we adopt the linear response framework, in which a small voltage bias \(\Delta V\) and a temperature difference \(\Delta T\) are applied between the left and right contacts in Fig.~\ref{Fig.sketch}. In this regime, the charge current \(I_e\) and heat current \(I_Q\) are given by
\begin{eqnarray}
I_e &=& - e^2 L_0 \Delta V + \frac{e}{T}\, L_1 \Delta T\ , \label{Ie} \\
I_Q &=& e L_1 \Delta V - \frac{1}{T}\, L_2 \Delta T\ , \label{Iq}
\end{eqnarray}
where \(T\) is the absolute temperature, and \(L_n\), \(n \in   \{ 0, 1, 2\} \), are the  thermal integrals defined by
\begin{equation}\label{thermal_integrals}
L_n(\mu, T) = \frac{2}{h} \int \limits_{-\infty}^{\infty} d\varepsilon\, \mathcal{T}(\varepsilon)\, (\varepsilon - \mu)^n \left( -\frac{\partial f(\varepsilon,\mu,T)}{\partial \varepsilon} \right),
\end{equation}
with \(\mathcal{T}(\varepsilon)\) the energy-dependent transmission function given by Eq. \ref{eq:Transmission}, \(h\) the Planck's constant, and \(f(\varepsilon,\mu,T)\) the Fermi–Dirac distribution function, which depends explicitly on energy $\varepsilon$, chemical potential $\mu$, and temperature $T$:
\begin{equation}
f(\varepsilon,\mu,T) = \frac{1}{2} \left(1 - \tanh\left( \frac{\varepsilon - \mu}{2k_B T} \right) \right).
\label{fermifunction}
\end{equation}

The Seebeck coefficient \(\mathcal{S}\) quantifies the voltage response to a temperature gradient under open-circuit conditions (\(I_e = 0\)). In the linear response limit, valid when \(|\Delta T| \ll T\) and \(|e \Delta V| \ll \mu\), the Seebeck coefficient \(\mathcal{S}\) is defined as
\begin{equation}
\mathcal{S} = \left. \frac{\Delta V}{\Delta T} \right|_{\displaystyle \tiny  I_e = 0},
\end{equation}
which corresponds to the voltage required to cancel the charge current under an applied temperature gradient. Substituting Eq.~\ref{Ie} into this condition yields
\begin{equation}
\mathcal{S}(\mu,T) = \frac{1}{e T} \frac{L_1}{L_0}.
\label{Seebeck}
\end{equation}

%
%

\begin{figure}[tb]
    \centering
       \includegraphics[width=\columnwidth]{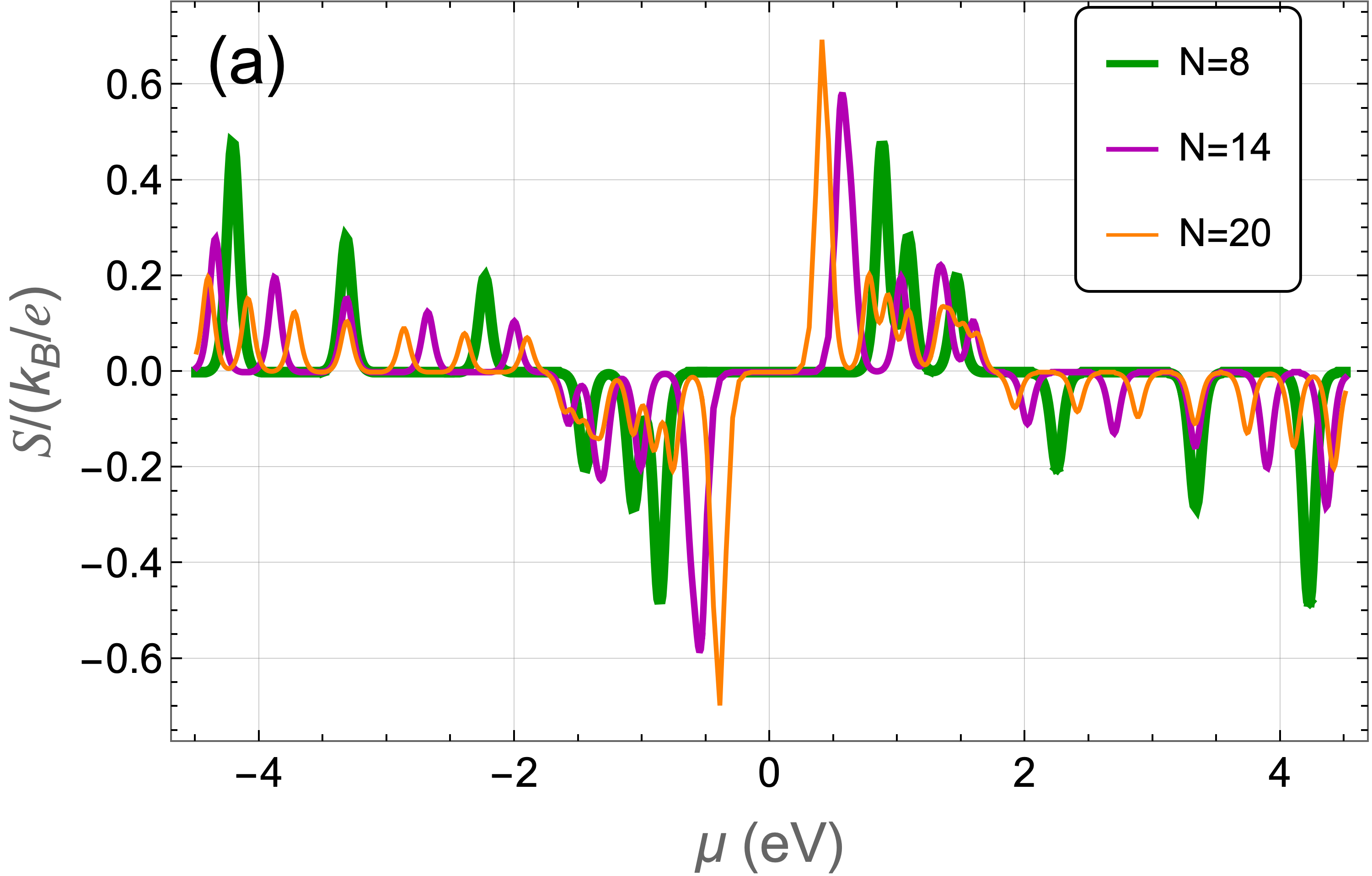}\\[4pt]
    \includegraphics[width=\columnwidth]{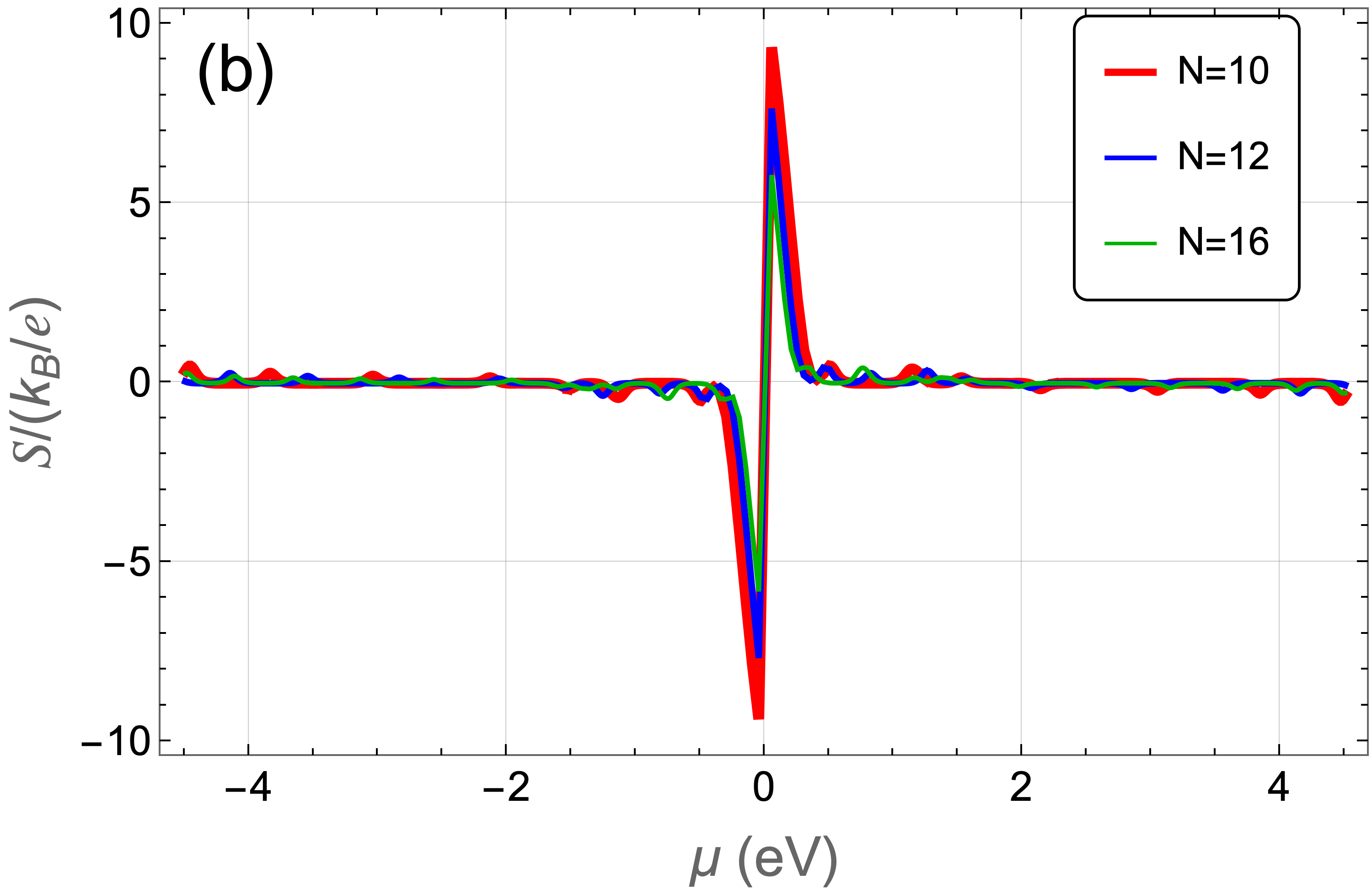}
    \caption{Seebeck coefficient $ \mathcal{S}$ as a function of chemical potential $\mu$ for armchair silicene nanoribbons at $T = 300$\,K. Panel (a) shows metallic ribbons with $N = 8$, $14$, and $20$, and panel (b) displays semiconducting ribbons with $N = 10$, $12$, and $16$.}
    \label{Fig.Seebeck}
\end{figure}

By using the thermal integrals of Eq.~\ref{thermal_integrals} into Eq.~\ref{Seebeck}, 
the Seebeck coefficient reads
\begin{equation}
\label{Seebeckin}
 \mathcal{S} =\frac{k_B}{e}\displaystyle \frac{\displaystyle \int \limits_{-\infty}^\infty \mathcal{T}(\varepsilon)\alpha(\varepsilon)\text{sech}^{2}\Big(\frac{\alpha(\varepsilon)}{2}\Big)d\varepsilon }{\displaystyle \displaystyle \int \limits_{-\infty}^{\infty} \mathcal{T}(\varepsilon)\text{sech}^{2}\Big(\frac{\alpha(\varepsilon)}{2}\Big)d\varepsilon}\;,
 \end{equation}
where $\alpha(\varepsilon)=(\varepsilon-\mu)/k_B T$, with $k_B$ the Boltzmann constant. This expression highlights how \(\mathcal{S}\) is governed by the energy dependence of the transmission function \(\mathcal{T}(\varepsilon)\), weighted by the thermal broadening encoded in the $\text{sech}^{2}\alpha(\varepsilon)/2$ function. 

To analyze the Seebeck coefficient $\mathcal{S}$, we apply Eq.\ (\ref{Seebeckin}) to metallic and semiconducting ASiNR with different widths, as shown in panels (a) and (b) of Fig.~\ref{Fig.Seebeck} respectively. Panel (a) presents the computed Seebeck coefficient as a function of \(\mu\) for metallic ribbons with widths \(N = 8\), \(14\), and \(20\), all at room temperature. In these metallic systems, $\mathcal{S}$ vanishes around the energy region where the electronic transmission $\mathcal{T}=1$, see Fig.~\ref{Fig.TDB} bottom-left panel for the ribbon $N=8$. Away from the first transmission channel $\mathcal{T}=1$ energy region, the curves exhibit a sequence of peaks and dips (\(|\mathcal{S}| \lesssim 0.7\,k_B/e\)) whose amplitude and position across $\mu$ depends on the ribbon width \(N\). For instance, in the case \(N = 8\), the first peaks occur near \(\mu \approx \pm 0.87\,\mathrm{eV}\), but their magnitude is smaller than that of wider metallic ribbons. This behavior reflects the underlying electronic band structure, as seen in Fig.~\ref{Fig.TDB} (right-bottom panel), where near the onset of parabolic and inverted bands the $\mathcal{S}$ peaks start to appear.    

In contrast, Fig.~\ref{Fig.Seebeck}(b) presents results for semiconducting ribbons with widths \(N = 10\), \(12\), and \(16\), corresponding to the \(N = 3p\) and \(N = 3p + 1\) armchair families. These ribbons exhibit pronounced dip–peak structures centered at \(\mu = 0\), with peak amplitudes reaching \(\mathcal{S} \approx \pm(7\text{–}10)\,k_B/e\), depending on the ribbon width. These large Seebeck responses observed in semiconducting ribbons arise from the presence of an energy gap combined with temperature. Electrons and holes contribute with the same magnitude but opposite signs with a peak-dip curve in $\mathcal{S}$ inside the band gap energy region. At the gap midpoint ($\mu = 0$), electrons and holes cancel each other, yielding $\mathcal{S}=0$ for all ribbon widths.  
As the ribbon width increases, the band gap narrows, and the 
peak-dip amplitudes decrease inside the band gap energy region.
These results emphasize the thermoelectric advantage of semiconducting nanoribbons, where the presence of larger band gaps as the ribbon width decreases enables substantially larger Seebeck coefficients than in gapless metallic systems.


\section{\label{sec:level4} Differential Entropy per Particle (DEP)}

The differential entropy per particle, denoted \(s(\mu, T)\), is defined thermodynamically as the partial derivative of the entropy with respect to particle number at constant temperature and volume, Eq.\ (\ref{eq:Sp_def}). Using standard thermodynamic identities and the structure of the Jacobian determinant, one obtains
\begin{equation}
\label{eq:Jacobian}
s(\mu, T) \equiv -\left(\frac{\partial \mu}{\partial T}\right)_N = 
\frac{\left(\partial N / \partial T\right)_\mu}{\left(\partial N / \partial \mu\right)_T}.
\end{equation}
This expression follows from the general identity that the product of cyclic partial derivatives over a closed variable set satisfies  
\((\partial \mu / \partial T)_N\,(\partial N / \partial \mu)_T\,(\partial T / \partial N)_\mu = -1\).

To evaluate \(s\) in practice, one needs an explicit expression for the number of particles $N$ as a function of \(\mu\) and \(T\). This can be written as
\begin{equation}
N(\mu,T) = \int \limits_{-\infty}^{\infty} d\varepsilon\, D(\varepsilon)\, f(\varepsilon,\mu,T),
\end{equation}
where \(D(\varepsilon)\) is the electronic density of states, Eq.~(\ref{eq:DOS}), and \(f(\varepsilon,\mu,T)\) is the Fermi–Dirac distribution function given by Eq.~(\ref{fermifunction}).

Since \(D(\varepsilon)\) is temperature independent, the temperature derivative of the number of particles in Eq.~(\ref{eq:Jacobian}) acts only on the Fermi function, leading to
\begin{equation}
\label{eq:S_sech}
 s=k_B\frac{\displaystyle \int \limits_{-\infty}^{\infty} D(\varepsilon)\alpha(\varepsilon)\text{sech}^{2}\Big(\frac{\alpha(\varepsilon)}{2}\Big)d\varepsilon }{\displaystyle \int \limits_{-\infty}^{\infty}D(\varepsilon)\text{sech}^{2}\Big(\frac{\alpha(\varepsilon)}{2}\Big)d\varepsilon}.
\end{equation}
In the high-temperature limit, where \(k_B T\) greatly exceeds the characteristic energy scale of the spectrum, the weighting function 
$\text{sech}^{2}\alpha(\varepsilon)/2$ becomes near constant over the energy integration range. Under this approximation, the integrand in the numerator becomes an odd function centered at \(\varepsilon = \mu\), causing the integral to vanish by symmetry. As a result, the differential entropy per particle approaches the limiting form
\begin{equation}
\label{eq:S_HT}
s(\mu,T) \simeq -\frac{\mu}{T},
\end{equation}
which differs from the Mott–Heikes formula for the Seebeck coefficient $\mathcal{S}$ [Eq.\ (\ref{eq:MottHeikes})] only by the factor \(1/e\). The direct comparison between Eq.~(\ref{Seebeckin}) and Eq.~\eqref{eq:S_sech}, reveals that \(\mathcal{S}\) and \(s\) share the same mathematical structure, but differ in the spectral function they depend on: while the Seebeck coefficient samples the transmission function \(\mathcal{T}(\varepsilon)\), the differential entropy per particle is governed by the density of states \(D(\varepsilon)\). The two quantities become equivalent \(\mathcal{T}(\varepsilon) \propto D(\varepsilon)\) when deep inside a clean band gap, where both functions are suppressed~\cite{cortes2023entropy}. We examine the correspondence between $s$ and $\mathcal{S}$ for ASiNRs in detail in the following sections. 

\begin{figure}[tb]
    \centering
     \includegraphics[width=\columnwidth]{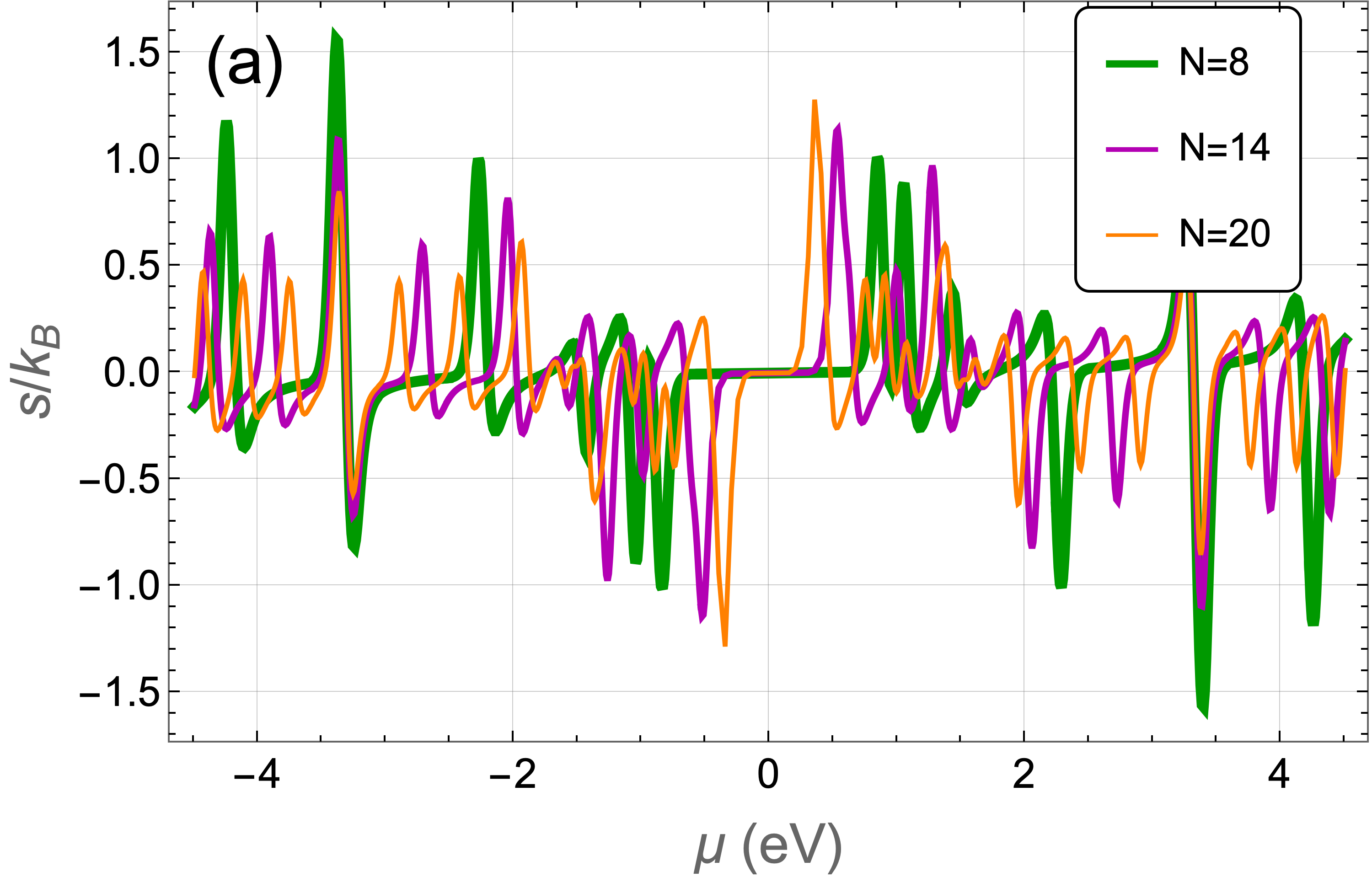}\\[4pt]
    \includegraphics[width=\columnwidth]{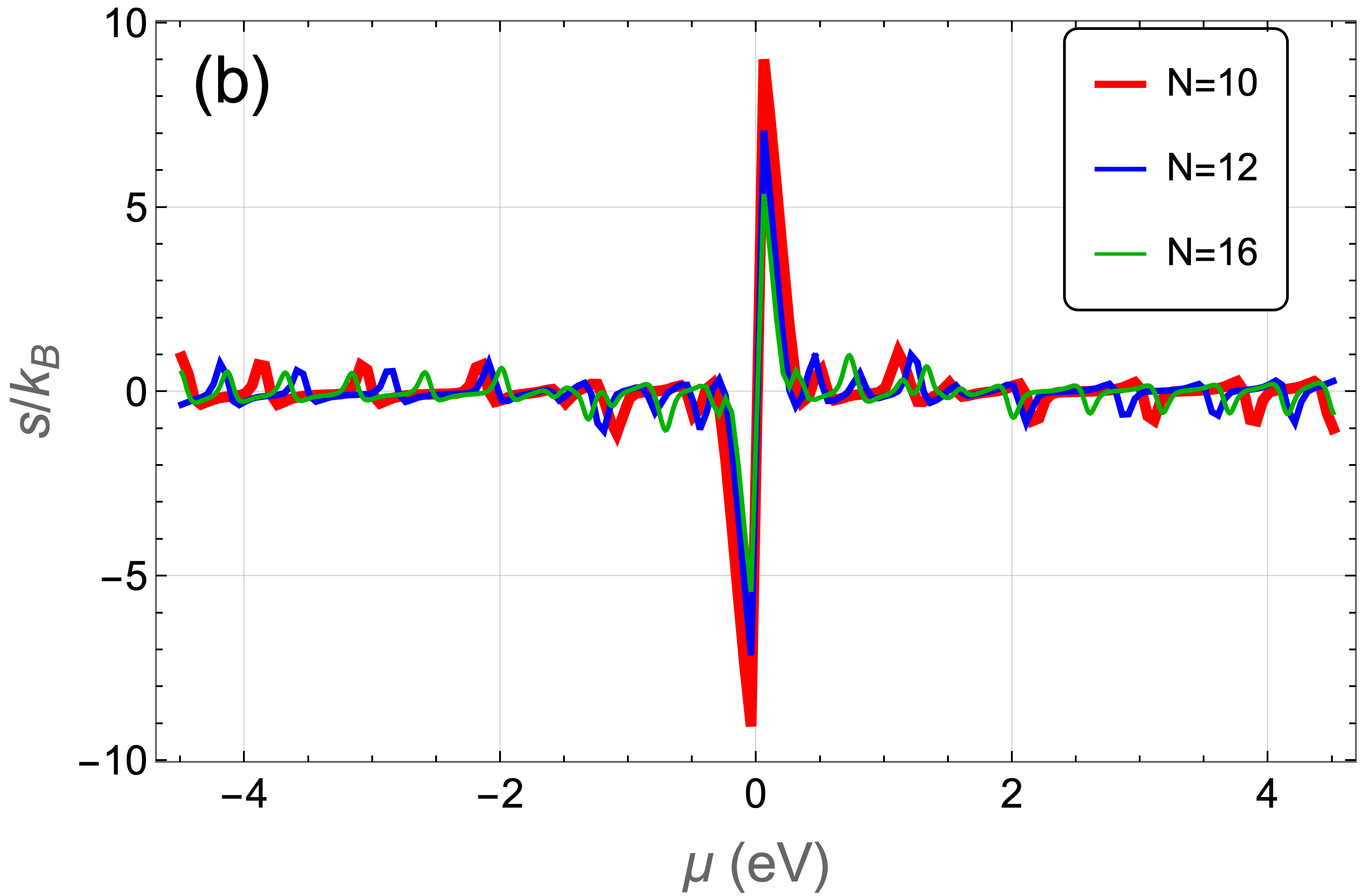}
    \caption{Differential entropy per particle $s$ as a function of chemical potential $\mu$ for armchair silicene nanoribbons at $T = 300\,\mathrm{K}$. Panel (a) shows metallic ribbons with widths $N = 8$, $14$, and $20$, while panel (b) corresponds to semiconducting ribbons with $N = 10$, $12$, and $16$.}
    \label{Fig.DEP}
\end{figure}

We now analyze the behavior of \(s(\mu,T)\) for the same ribbon widths we have used for the Seebeck coefficient in Fig.\ \ref{Fig.Seebeck}. 
Figure \ref{Fig.DEP} panel (a) shows the calculated differential entropy per particle $s$ as a function of chemical potential \(\mu\) for the metallic ribbons, and panel (b) for the semiconducting ones, all ribbons at \(T = 300\,\mathrm{K}\). The metallic ribbons in Fig.~\ref{Fig.DEP}(a) with widths \(N = 8\), \(14\), and \(20\), exhibit a linear DEP about $\mu=0$. This $\mu$ energy region with linear DEP response corresponds to a finite parabolic DOS, see Fig.~\ref{Fig.DOS12}(a) for $N=8$. We have checked that metallic ribbons with larger $N$ show the same parabolic DOS behavior. Away from charge neutrality ($\varepsilon=0$), the DEP displays a sequence of asymmetric peaks and dips with amplitudes \(|s| \lesssim 1.5\,k_B\), each one of them associated with the crossing of \(\mu\) through a van Hove (vH) singularity. As the ribbon width increases, the DEP shows more dips and peaks across $\mu$, associated with more available states and vH singularities in the DOS.

Figure ~\ref{Fig.DEP}(b) displays results for semiconducting ribbons (\(N = 10\), \(12\), and \(16\)), which belong to the \(N = 3p\) and \(N = 3p + 1\) families. These systems exhibit pronounced dip–peak structures centered at \(\mu = 0\), with extrema reaching \(\pm(5\text{–}10)\,k_B\). The magnitude of these peaks and the energy region where they are at, decreases with increasing $N$. Outside the band gap region, the behavior for the DEP oscillates in connection with the van Hove singularities of the DOS.

As one can notice, the results for the Seebeck coefficient and DEP for energy regions around $\mu=0$ are especially interesting. While for metallic ASiNRs we find finite DOS and transmission, for semiconducting ribbons the DOS and transmission are nearly zero. These different behaviors manifest in evident differences and similitude between $\mathcal{S}$ and $s$, as we examine quantitatively in the next section.

\section{Comparison between Seebeck and DEP}
\label{sec:SEE-DEP}

\begin{figure}[tb]
    \centering
      \includegraphics[width=\columnwidth]{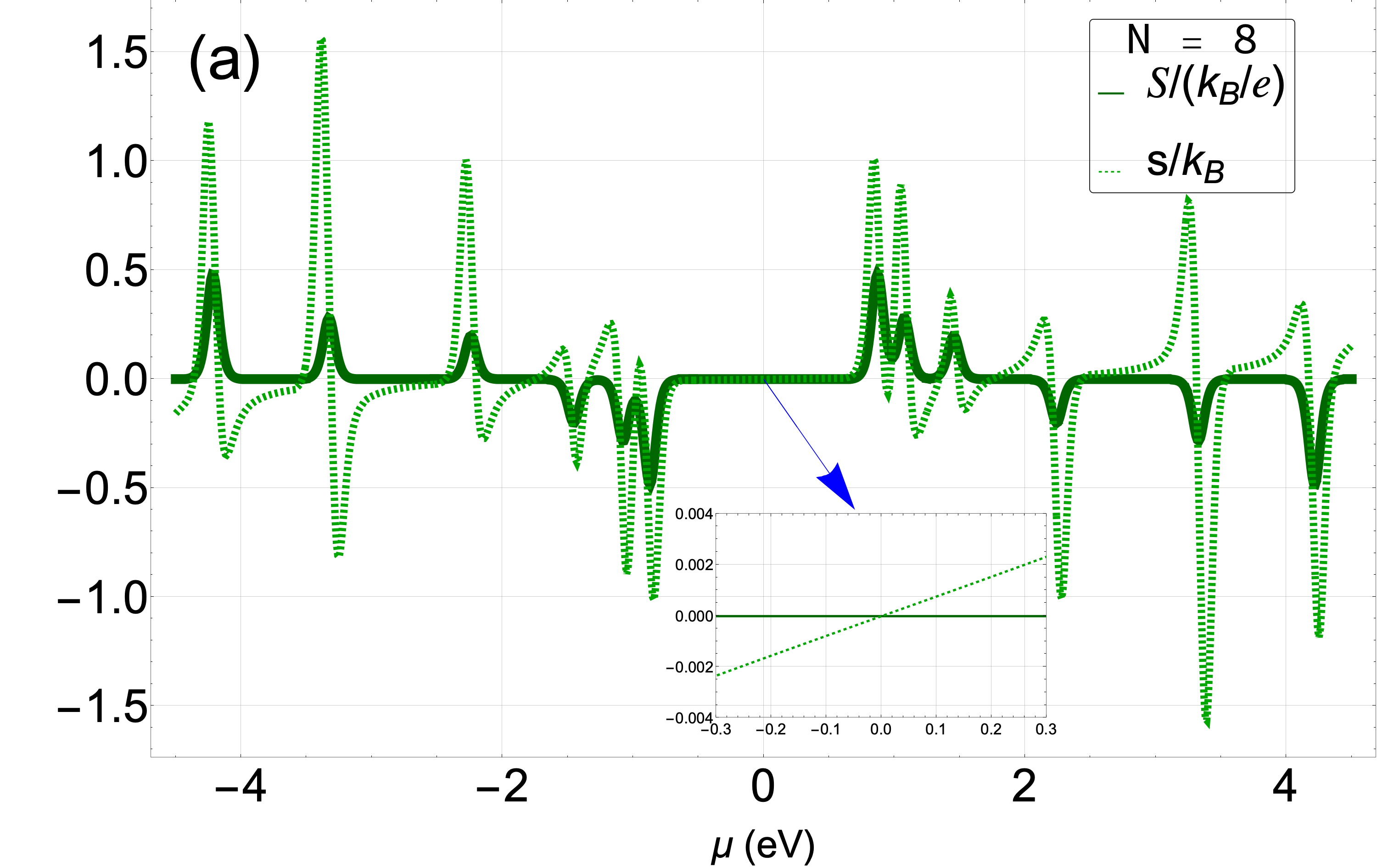}\\[4pt]
    \includegraphics[width=\columnwidth]{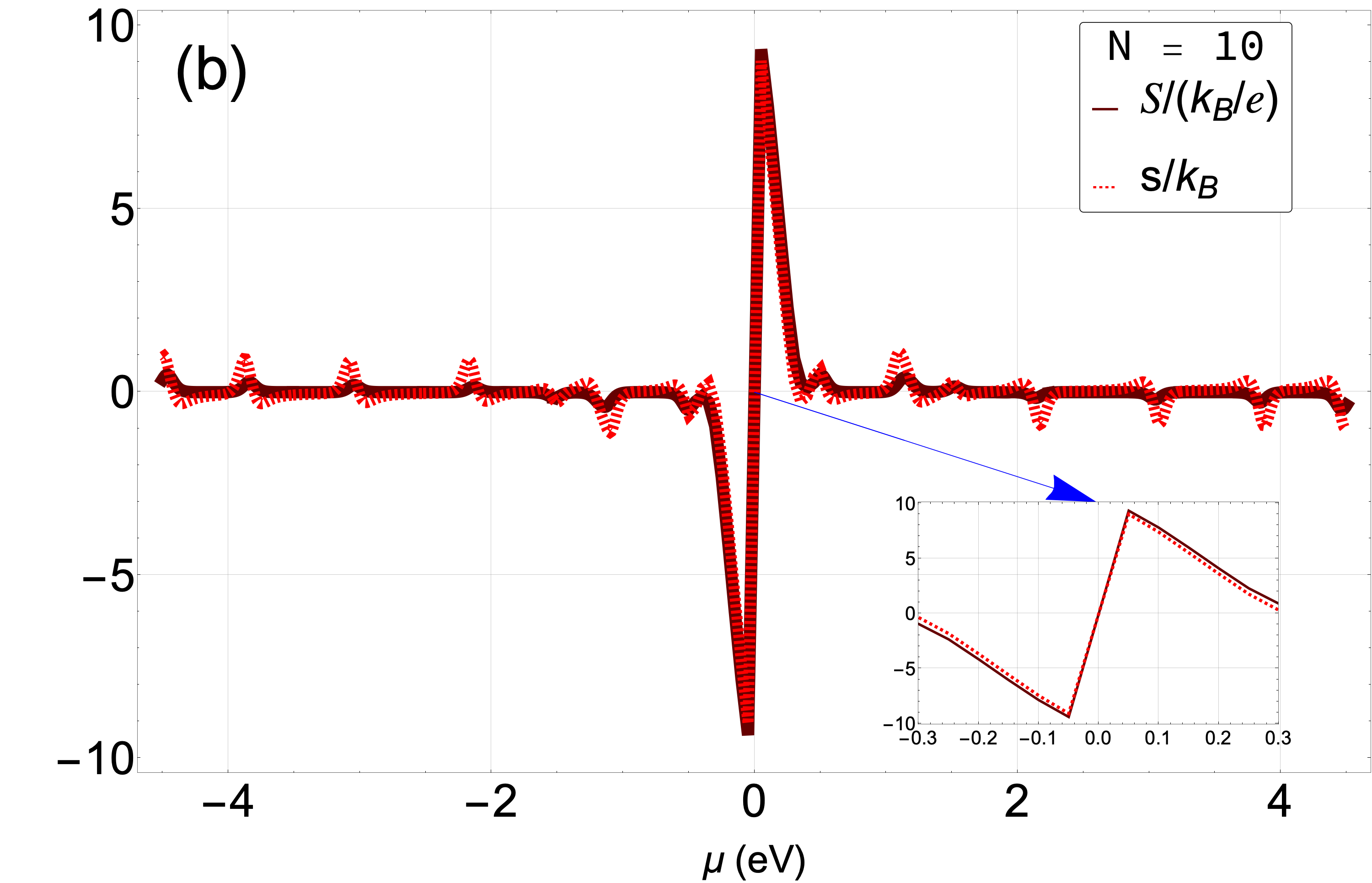}

    \caption{Comparison between the Seebeck coefficient $\mathcal{S}/(k_B/e)$ and the differential entropy per particle $s/k_B$ as a function of chemical potential $\mu$, at $T=300$ K. Panel (a) shows a metallic nanoribbon with $N = 8$, and panel (b) a semiconducting one with $N = 10$. The insets provide zoomed-in views of the central region around \(\mu = 0\).}
    \label{Fig.SEE-DEP}
\end{figure}

To evaluate the correspondence between the Seebeck coefficient \(\mathcal{S}\) and the differential entropy per particle \(s\), we compare their behavior for representative ribbon widths at \(T = 300\,\mathrm{K}\), as shown in Fig.~\ref{Fig.SEE-DEP}. Panel (a) corresponds to a metallic ribbon \(N = 8\), and panel (b) presents a semiconducting one \(N = 10\). In both panels, solid lines denote \(\mathcal{S}/(k_B/e)\), and dashed lines \(s/k_B\); the insets zoom in on the central region near \(\mu = 0\), highlighting key similitude and differences between \(s\) and \(\mathcal{S}\), as the level of agreement within the gap for the semiconducting case, and deviations in the metallic one.

\begin{figure}[tb]
    \centering
     \includegraphics[width=\columnwidth]{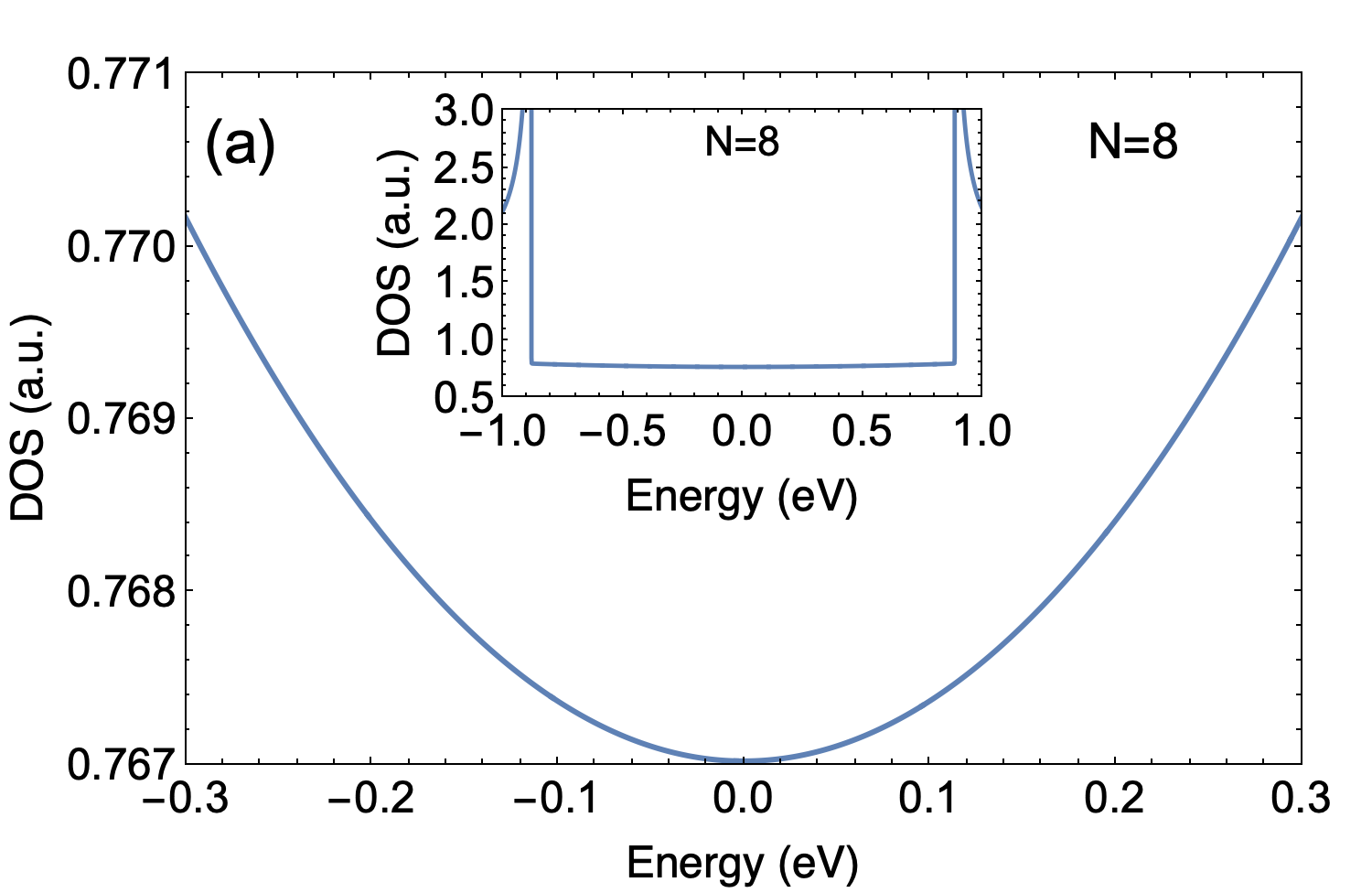}\\[4pt]
    \includegraphics[width=\columnwidth]{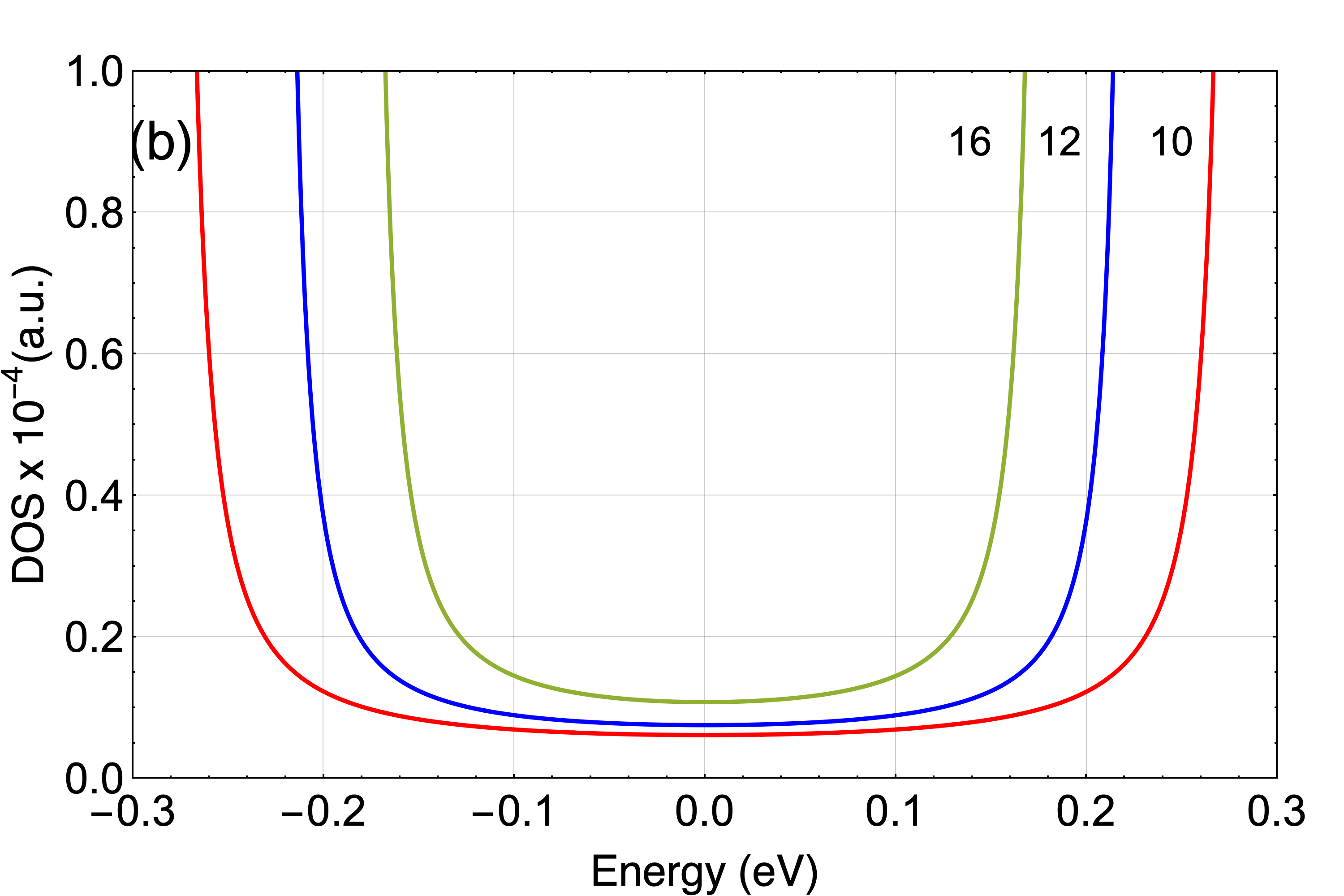}
    \caption{
(a) Density of states (DOS) in arbitrary units as a function of energy for the metallic nanoribbon with width $N = 8$. The DOS is not constant and exhibits slight variations in the energy region $|\varepsilon| \le 0.3$\,eV. The inset shows the DOS for the same nanoribbon $N=8$ as in Fig.\ \ref{Fig.TDB} (red lines) in the energy windows $|\varepsilon| \le 1$\,eV.
(b) DOS as a function of energy for semiconducting nanoribbons with widths $N = 10$, $N = 12$, and $N = 16$. Although the DOS remains small in the vicinity of $\varepsilon = 0$ eV, it is not strictly zero within this energy window.}
    \label{Fig.DOS12}
\end{figure}

For metallic ribbons in Fig.~\ref{Fig.SEE-DEP}(a), \(s\) and \(\mathcal{S}\) display a similar pattern of extrema across $\mu$ and away from $\mu=0$, but their amplitudes and energy alignments differ significantly. When both quantities approach $\mu=0$, they show different linear responses, while $s$ present a linear dependence on $\mu$ with a positive slope, $\mathcal{S}=0$ for the same $\mu$ region (see inset of Fig.\ \ref{Fig.SEE-DEP}(a)) up to the first peaks appear in both quantities. In the energy window $\mu \simeq \pm 0.9$ around $\mu=0$, the transmission function \(\mathcal{T}(\varepsilon)\), and the density of states \(D(\varepsilon)\) remain finite, see bottom panels in Fig.\ \ref{Fig.TDB}. However, $D(\varepsilon)$ is not strictly constant; instead it exhibits a parabolic dependence about energy 
$\varepsilon=0$, as shown in Fig.\ \ref{Fig.DOS12}(a). As a consequence, the DEP acquires a finite positive slope around $\mu =0$, leading to the ratio $s/\mathcal{S}$ undefined for that $\mu$ region, and $s/\mathcal{S}=0/0$ at $\mu=0$.

By contrast, the semiconducting ribbon in Fig.~\ref{Fig.SEE-DEP}(b), exhibits a prominent dip–peak structure centered at \(\mu = 0\), where both \(s\) and \(\mathcal{S}\) show high similitude within the energy gap region, and reaching their extrema with excellent agreement: \(|\mathcal{S}| \approx 10\,k_B/e\) and \(|s| \approx 10\,k_B\). This alignment originates from the presence of an electronic gap, where \(\mathcal{T}(\varepsilon)\) and \(D(\varepsilon)\) nearly vanish, confining the integrals of Eqs.~\eqref{Seebeckin} and~\eqref{eq:S_sech} to a narrow energy region of equivalence. 
We have checked for the semiconducting ribbon that $\mathcal{T}(\varepsilon)$ and $D(\varepsilon)$ present a flat line shape, although they are not strictly zero in the band gap energy region as expected for a semiconductor, due to the Green's function formalism in Eq.\ \ref{eq:DOS}. We show this behavior only for the DOS in Fig.\ \ref{Fig.DOS12}(b).  
Therefore, for semiconducting armchair ribbon systems with $N=3p$ or $N=3p+1$ width condition ($p \in \mathbb{Z}>0$), we anticipate that the ratio $s/\mathcal{S} \approx e$ holds, arising from a spectral \textit{coincidence} that aligns transport and thermodynamic quantities within the ribbon gapped regime, as also seen in gapped zigzag graphene ribbons \cite{cortes2023entropy}.

\section{DEP and Seebeck as a tool for quantifying the electron charge}

\begin{figure}[tb]
    \centering
      \includegraphics[width=\columnwidth]{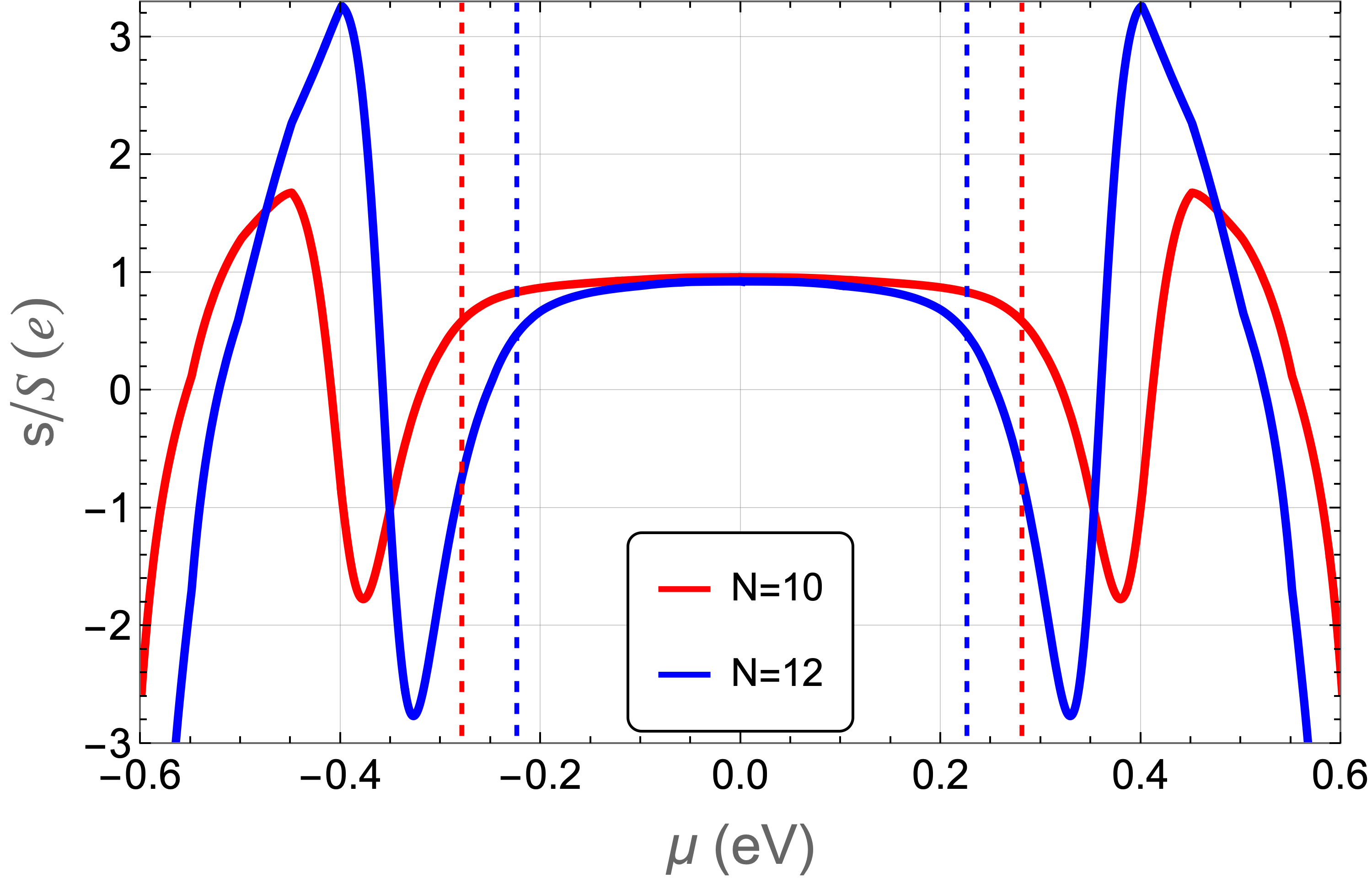}\\[4pt]
    \includegraphics[width=\columnwidth]{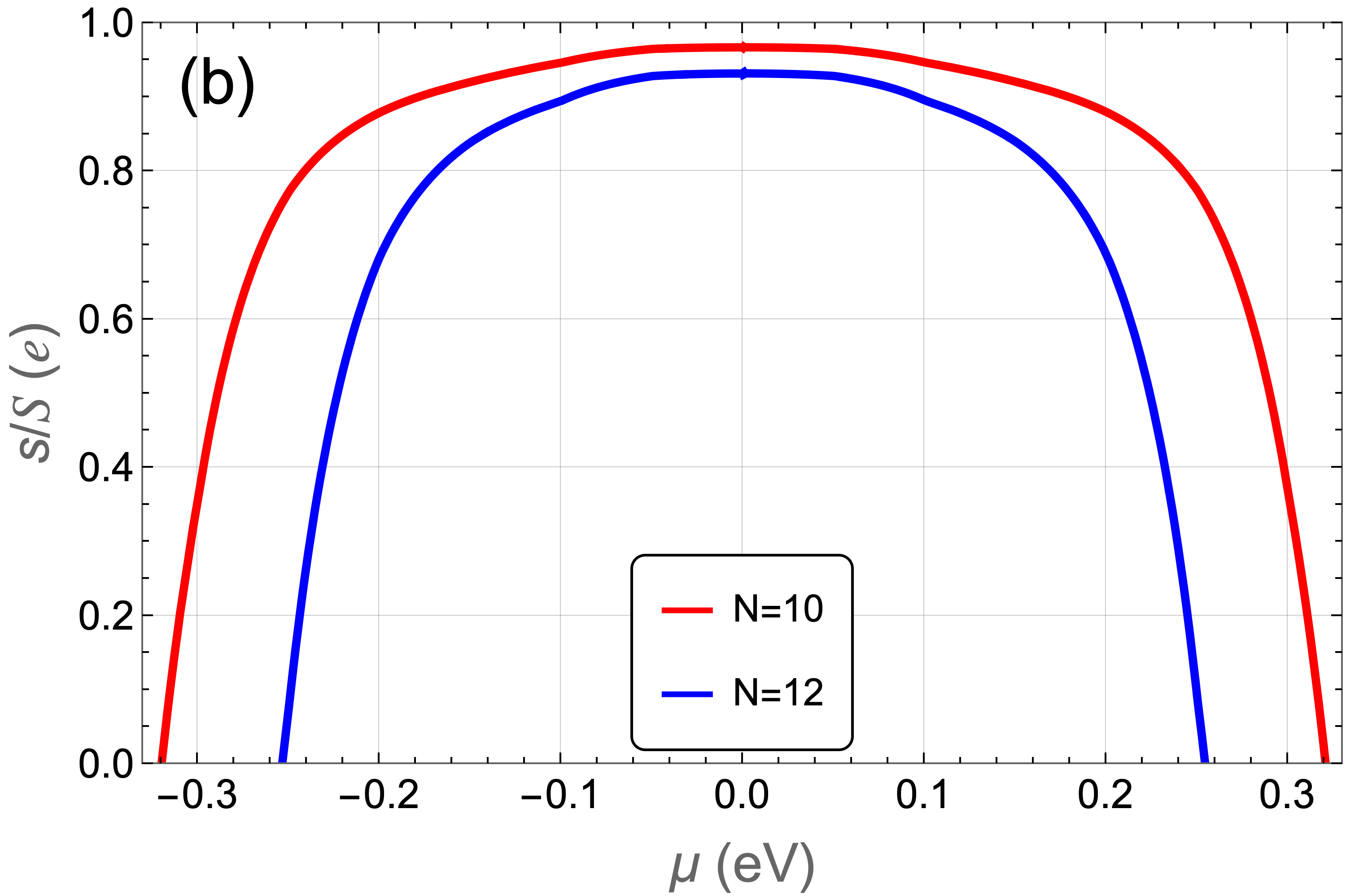}
    \caption{ 
    (a) Ratio \(s/\mathcal{S}\) in units of the elementary charge $e$ as a function of the chemical potential \(\mu\) for nanoribbons with \(N=10\) (solid red line) and \(N=12\) (solid blue line) at \(T = 300\,\mathrm{K}\). 
    In both cases, the ratio approaches unity as \(\mu \rightarrow 0\). The red dashed vertical lines indicate the energy gap edges for \(N=10\), located at \(\mu \simeq \pm 0.28\) eV, and blue dashed vertical lines denote the gap edges for \(N=12\), located at \(\mu \simeq \pm 0.22\) eV.  (b) Zoomed-in view of panel (a) around $\mu = 0$, highlighting the behavior of $s/\mathcal{S}$.}
    \label{fig:SoverS_N10}
\end{figure}

In previous Section \ref{sec:SEE-DEP}, we established the conditions under which the Seebeck coefficient \( \mathcal{S} \) and the differential entropy per particle \( s \) are highly equivalent: the ribbon width corresponds to $N=3p$ or $N=3p+1$, with $p \in \mathbb{Z}>0$. This ensures there are band gaps for the spectrum of both the transmission function $\mathcal{T}(\varepsilon)$, and the DOS $D(\varepsilon)$, with their amplitudes nearly zero at the same energy range. Knowing that information, we examine the ratio \( s/\mathcal{S} \) for two ribbon widths $N$ fulfilling that condition, $N=10$ and $N=12$, both as a function of chemical potential $\mu$ to assess its reliability for estimating the elementary electron charge at room temperature.

Figure~\ref{fig:SoverS_N10} shows the ratio \( s/\mathcal{S} \) in units of \( e \) for a $\mu$ range near the band gap region for the two ribbon widths \( N = 10 \) (solid red lines), and \( N = 12 \) (solid blue lines), both ribbons at \( T = 300\,\mathrm{K} \). The ratio $s/\mathcal{S}$ is mirror symmetric about $\mu=0$ eV for both ribbon widths as the band structure presents electron-hole symmetry about $\varepsilon=0$ eV. The ratio exhibits symmetric positive peaks near $\mu \simeq \pm0.44$ eV ($N=10$), $\mu \simeq \pm0.4$ eV ($N=12$), associated with a positive ratio of the first oscillations appearing just after the dip-peak curves for $s$ and $\mathcal{S}$, see Fig. \ \ref{Fig.SEE-DEP}(b). As $\mu$ approaches the conduction and valence band edges, symmetric negative minima appear due to the opposite signs for $s$ and $\mathcal{S}$. As $\mu$ goes to zero, the ratio $s/\mathcal{S}$ starts to increase, and a nearly flat plateau is reached for both ribbon widths. This flat behavior for $s/\mathcal{S}$ spans within a magnitude of $|\mu| \simeq 0.30$ eV ($N=10$), and $|\mu| \approx 0.1$ eV ($N=12$). Within this $\mu$ energy range, the ratio $s/\mathcal{S} \approx e$ holds, providing strong evidence that the DEP, $s$, serves as a direct thermodynamic analog of the Seebeck coefficient $\mathcal{S}$ \cite{peterson2010kelvin,goupil2011thermodynamics} in the presence of a clean electronic gap, and $\mathcal{S} \simeq s/e$ characterizes the transported entropy per charge at room temperature, confirming a thermodynamic–transport correspondence for gapped silicene ribbons.

\begin{figure}[tb]
    \centering
      \includegraphics[width=\columnwidth]{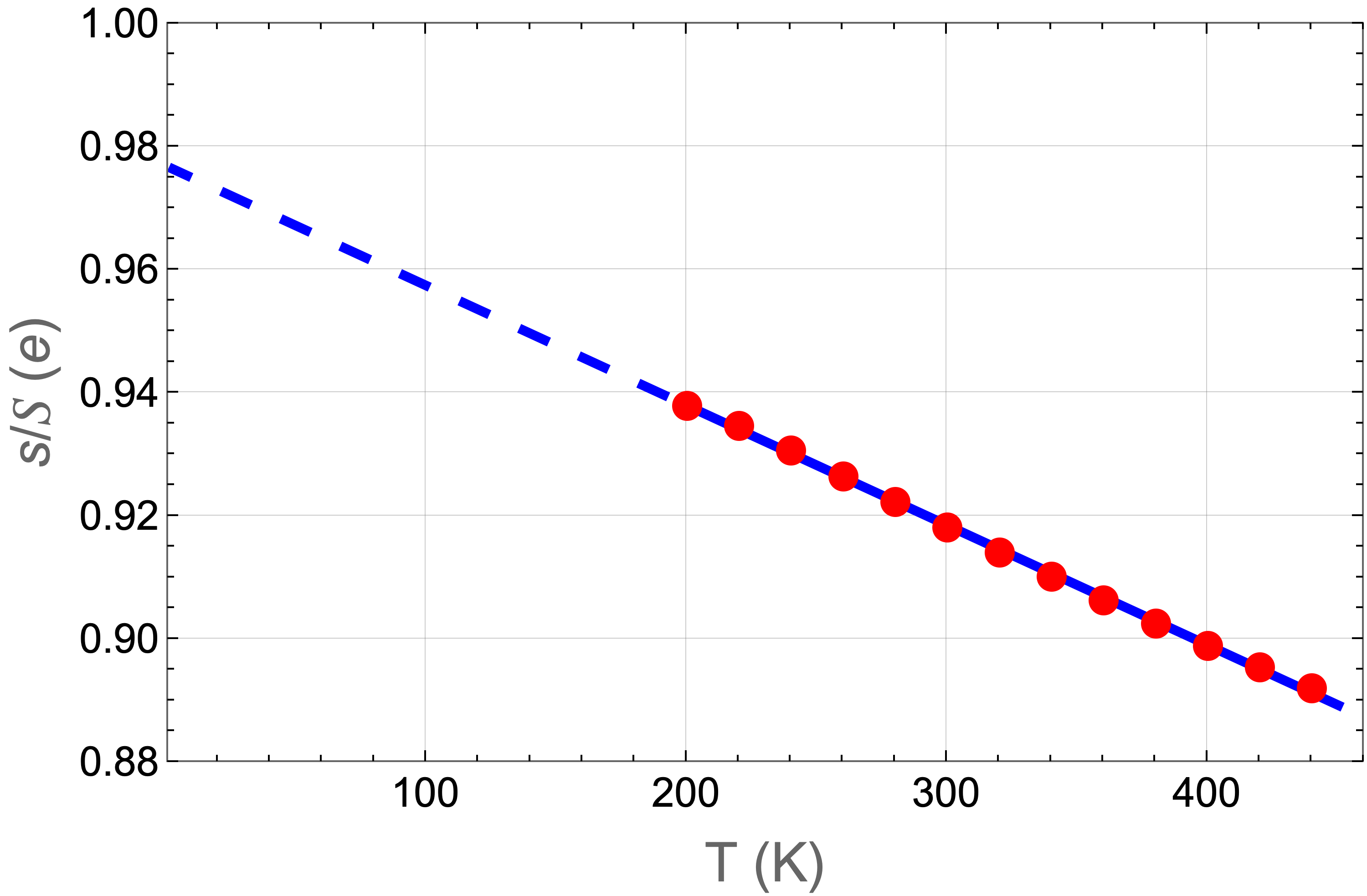}\\[4pt]
    \caption{
The ratio $s/\mathcal{S}$ expressed in units of the elementary charge $e$, is shown as a function of temperature $T$ for a semiconductor nanoribbon with width $N = 10$ [see Fig.\ \ref{fig:SoverS_N10}(b)]. To ensure numerical stability within the energy gap—where the density of states is nearly zero—we selected a chemical potential of $\mu = -0.15 \, \text{eV}$. Numerical computations at $T < 200\,\text{K}$ proved to be highly unstable. The solid line represents a linear fit to the data, from which the estimated value of the elementary charge is found to be approximately 2\% lower than the real value.
 }
    \label{Fig.RazonsST}
\end{figure}

This analysis indicates that a flat profile in \( s/\mathcal{S} \) identifies spectral regions where thermodynamic and transport quantities are governed by the same set of gapped energy states. We also observe that at a fixed temperature of $T = 300$ K, the reduction of the ribbon width—and hence the number of active conducting armchair channels—broadens the $\mu$ range over which the ratio \( s/\mathcal{S} \approx e \) holds. Finally, we have calculated the ratio $s/\mathcal{S}$ for the ribbon width $N=10$ at a fixed chemical potential of $\mu=-0.15$ eV and different temperatures. As shown in Fig.\ \ref{Fig.RazonsST}, the ratio linearly decreases with temperature, deviating from its perfect value of $1e$ as $T$ increases. These behaviors for the ratio $s/\mathcal{S}$ show that it is also possible to estimate charge fluctuations present in gapped 2D materials at finite temperatures.     

\section{Conclusions}
\label{sec:Conclusions}

We have demonstrated a quantitative correspondence between transport and thermodynamic descriptions of charge carriers in armchair silicene nanoribbons using a \(\pi\)-orbital tight-binding model combined with nonequilibrium Green’s function techniques, allowing us to obtain the Seebeck coefficient \(\mathcal{S}\) and the differential entropy per particle  \(s\) for both metallic and semiconducting regimes at/near room temperature. We systematically compared $\mathcal{S}$ and $s$, and calculated their ratio $s/\mathcal{S}$ for different ribbon widths $N$.

In metallic ribbons, we find significant discrepancies between $\mathcal{S}$ and $s$ as a function of chemical potential $\mu$, and the ratio $s/\mathcal{S}$ is undetermined for these ribbon systems. However, in semiconducting ribbons, \(\mathcal{S}\) and \(s\) display nearly identical dip–peak structures within the band gap energy region, with amplitudes reaching approximately \(10\,k_B\) and \(10\,k_B/e\), respectively, for a ribbon with width $N=10$. This close agreement originates from the presence of a clean band gap, where both the transmission function \(\mathcal{T}(\varepsilon)\) and the density of states \(D(\varepsilon)\) nearly vanish, thereby restricting the energy integrals that define \(s\) and \(\mathcal{S}\) to the same $\mu$-dependent gapped window. As a result, their ratio remains nearly constant for the band gap region as \(s/\mathcal{S}\approx e\). While this behavior is reminiscent of Kelvin’s thermodynamic formulation of the Seebeck coefficient, the ratio $s/\mathcal{S}\approx e$ is a consequence of specific spectral conditions in gapped systems that perfectly align the transport and thermodynamic observables. This provides a practical route to estimate the elementary charge and gain insight into how clean a band gap is when the ratio shows a value of $e$.

Taken together, these results establish \(s\) as a robust and purely thermodynamic proxy for the Seebeck coefficient in 2D systems with well-defined electronic gaps. Since this approach relies primarily on spectral properties rather than electron–electron interactions, it can be extended to a broad class of 2D and quasi-one-dimensional materials, offering the optimization of thermoelectric performance through the sensitivity of $s$, and a transport-thermodynamic framework for the elementary charge determination.

\begin{acknowledgments}
The authors acknowledge financial support from ANID Fondecyt grant no. 1250173 (F.J.P., P.V.) and 1240582 (F.J.P., N.C., P.V.). B.C acknowledge PUCV and UTFSM. B.C. acknowledges the support of ANID Becas/Doctorado Nacional 21250015. P.V. acknowledges the support of Cedenna grant no. CIA250002. 
\end{acknowledgments}
 
\bibliographystyle{apsrev4-1}
\bibliography{aapmsamp}

\end{document}